\documentclass[paper=a4,fontsize=12pt,DIV=12]{scrartcl}

\usepackage{lmodern}
\usepackage[utf8]{inputenc}
\usepackage[T1]{fontenc}
\usepackage{amsmath}
\usepackage{amssymb}
\usepackage{authblk}
\usepackage{bm}
\usepackage{bookmark}
\usepackage{booktabs}
\usepackage{caption}
\usepackage[nodayofweek]{datetime}
\usepackage{enumitem}
\setlist{labelindent=0pt}
\setlist{leftmargin=*}
\usepackage{graphicx}
\usepackage[labelformat=simple]{subcaption}

\usepackage{siunitx}
\usepackage{xcolor}
\usepackage{xspace}

\hypersetup{hidelinks}
\usepackage[capitalise]{cleveref}

\usepackage[mathlines]{lineno}%
\newcommand*\linenomathpatchAMS[1]{%
\expandafter\pretocmd\csname #1\endcsname {\linenomathAMS}{}{}%
\expandafter\pretocmd\csname #1*\endcsname{\linenomathAMS}{}{}%
\expandafter\apptocmd\csname end#1\endcsname {\endlinenomath}{}{}%
\expandafter\apptocmd\csname end#1*\endcsname{\endlinenomath}{}{}%
}
\expandafter\ifx\linenomath\linenomathWithnumbers
\let\linenomathAMS\linenomathWithnumbers
\patchcmd\linenomathAMS{\advance\postdisplaypenalty\linenopenalty}{}{}{}
\else
\let\linenomathAMS\linenomathNonumbers
\fi
\AtBeginDocument{%
\linenomathpatchAMS{gather}
\linenomathpatchAMS{multline}
\linenomathpatchAMS{align}
\linenomathpatchAMS{alignat}
\linenomathpatchAMS{flalign}
}

\usepackage[backend=biber,giveninits=true,doi=false,sorting=none,style=numeric-comp]{biblatex}
\DeclareFieldFormat[article]{citetitle}{\textit{#1}\isdot}
\DeclareFieldFormat[article]{title}{\textit{#1}\isdot}
\DeclareFieldFormat[unpublished]{citetitle}{\textit{#1}\isdot}
\DeclareFieldFormat[unpublished]{title}{\textit{#1}\isdot}
\DeclareFieldFormat[inproceedings]{citetitle}{\textit{#1}\isdot}
\DeclareFieldFormat[inproceedings]{title}{\textit{#1}\isdot}
\DeclareNameAlias{author}{family-given}

\addtokomafont{disposition}{\rmfamily}   
\addtokomafont{subsection}{\normalsize}
\addtokomafont{section}{\large}
\KOMAoptions{fontsize=10pt,twocolumn,DIV=20}

\usepackage{acronym}
\newacro{SM}{standard model}
\newacro{BSM}{beyond-the-SM}
\newacro{ML}{machine learning}
\newacro{LHC}{Large Hadron Collider}
\newacro{CNN}{convolutional neural network}
\newacro{QCD}{quantum chromodynamics}
\newacro{VAE}{variational autoencoder}
\newacro{SVDD}{support vector data description}
\newacro{CLR}{contrastive learning}
\newacro{HEP}{High Energy Physics}
\newacro{NF}{Normalizing Flow}
\newacro{MC}{Monte Carlo}
\newacro{ROC}{Receiver Operating Characteristic}
\newacro{2HDM}{two-Higgs-doublet model}
\newacro{SIMP}{Strongly Interacting Massive Particle}

\newcommand{\pT}{\ensuremath{p_{\mathrm{T}}}\xspace}
\newcommand{\pTindex}[1]{\ensuremath{p_{\mathrm{T,#1}}}\xspace}
\newcommand{\abseta}{\ensuremath{\left|\eta\right|}\xspace}
\newcommand{\tauone}{\ensuremath{\tau_1}\xspace}
\newcommand{\tautwo}{\ensuremath{\tau_2}\xspace}
\newcommand{\tauthree}{\ensuremath{\tau_3}\xspace}
\newcommand{\donetwo}{\ensuremath{\sqrt{d_{12}}}\xspace}
\newcommand{\dtwothree}{\ensuremath{\sqrt{d_{23}}}\xspace}
\newcommand{\ecftwo}{\ensuremath{\mathit{ECF}_2}\xspace}
\newcommand{\ecfthree}{\ensuremath{\mathit{ECF}_3}\xspace}

\title{RODEM Jet Datasets}

\author[1,2]{Knut Zoch}
\author[1]{John Andrew Raine}
\author[1]{Debajyoti Sengupta}
\author[1]{Tobias Golling}
\affil[1]{Département de physique nucléaire et corpusculaire, Université de Genève, 1211 Genève, Switzerland}
\affil[2]{Laboratory for Particle Physics and Cosmology, Harvard University, Cambridge, Massachusetts 02138, USA}
\date{}
\publishers{%
\normalfont\normalsize%
\parbox{0.9\linewidth}{%
    We present the \emph{RODEM Jet Datasets}, a comprehensive collection of simulated large-radius jets designed to support the development and evaluation of machine-learning algorithms in particle physics.
    These datasets encompass a diverse range of jet sources, including quark/gluon jets, jets from the decay of $W$~bosons, top quarks, and heavy new-physics particles.
    The datasets provide detailed substructure information, including jet kinematics, constituent kinematics, and track displacement details, enabling a wide range of applications in jet tagging, anomaly detection, and generative modelling.\\[0.2ex]
    
    The datasets are available on Zenodo: \href{https://doi.org/10.5281/zenodo.12793616}{10.5281/zenodo.12793616}
}
}

\begin{document}

\maketitle

\section{Introduction}
\label{sec:intro}

Hadron colliders, such as the \ac{LHC} at CERN, are at the forefront of exploring the fundamental constituents of matter and the forces governing their interactions.
These machines accelerate protons or heavy ions to near-light speeds and collide them, recreating conditions akin to those just moments after the Big Bang.
One of the most prolific results of these high-energy collisions is the production of jets -- collimated streams of particles resulting from the hadronisation of quarks and gluons.
Among these, large-radius jets are particularly interesting as they can arise not only from quarks and gluons but also from collimated decay products of heavier particles.
Their complex substructure encapsulates detailed information about the high-energy processes that generated them, making them crucial objects to study in searches for new physics.

Large-radius jets are not only pivotal for searches for new physics but also for precise \ac{SM} measurements. 
Their substructure composition provides valuable insights, particularly when advanced jet tagging techniques are used to differentiate between various jet types.
The interest in applying \ac{ML} to large-radius jets is substantial due to the numerous potential applications it offers: 
from enhancing jet tagging accuracy to building generative jet models and detecting anomalies that may indicate novel phenomena.
The versatility and efficacy of these \ac{ML} applications underscore their transformative impact on the study of large-radius jets in particle physics.

The successful application of \ac{ML} techniques to large-radius jets relies heavily on the availability of high-quality datasets of simulated jets.
These datasets are essential for training and validating \ac{ML} models, ensuring they can accurately capture the substructures and characteristics of jets produced in hadron-collider experiments.
To enable a wide range of applications, these datasets must provide both jet-level information, such as jet kinematics, and detailed substructure information at the jet constituent level, including the kinematics of the constituents, their track compatibility with the primary vertex, and more.
Several datasets have been developed to meet these needs, including the top quark tagging dataset~\cite{toptagging}, the \emph{JetNet} dataset~\cite{jetnet}, the \emph{JetClass} dataset~\cite{jetclass}, and others~\cite{quark-gluon-tagging,higgs-tagging,jedi-net}.
However, many provide limited substructure information or are only available for a restricted range of jet types.

This note presents the \emph{RODEM Jet Datasets}, which aim to complement existing datasets by providing comprehensive and fine-grained substructure information across a diverse range of jet types.%
\footnote{\emph{RODEM} -- ``Robust Deep Density Models for High-Energy Particle Physics and Solar Flare Analysis'', the SNSF Sinergia grant that partially funded this project.}
The datasets include simulated jets from proton-proton collisions at a centre-of-mass energy of $\sqrt{s} = \SI{13}{\TeV}$, reflecting \ac{LHC} Run~2 conditions.
Jet detection is simulated using a detector model resembling the ATLAS experiment~\cite{atlas-experiment} at the \ac{LHC}, with large-radius jet reconstruction following typical ATLAS standards.
The datasets encompass various jet sources, from single-prong quark/gluon jets to multi-prong jets from decays of $W$ bosons, top quarks, and heavy \ac{BSM} particles, such as heavy scalar bosons.
They provide full kinematic information and a range of substructure metrics at the jet level.
Additionally, the kinematics of up to one hundred constituents are detailed per jet.
For charged constituents, the datasets include charge information and track displacement details relative to the primary interaction vertex, as in the \emph{JetClass} dataset.

The datasets are available on \href{https://doi.org/10.5281/zenodo.12793616}{Zenodo}~\cite{dataset}.

\section{Datasets}\label{sec:data}

This section provides a detailed overview of the datasets generated for the \emph{RODEM Jet Datasets} project.
The aim is to offer a comprehensive jet dataset resource with an emphasis on fine-grained substructure information across a wide variety of jet types.
The \emph{RODEM Jet Datasets} are structured to support a broad spectrum of machine learning applications, from jet tagging to generative models and anomaly detection.

Particle kinematics are described using the usual \ac{LHC} coordinate system.
In the transverse plane, cylindrical coordinates~$(r,\phi)$ are used, with $\phi$~being the azimuthal angle around the \(z\)-axis, which follows the direction of the colliding beams.
The pseudorapidity~$\eta$ is defined in terms of the polar angle~$\theta$ as $\eta = -\ln \tan(\theta/2)$.
Angular distance is measured in units of $\Delta R \equiv \sqrt{(\Delta\eta)^{2} + (\Delta\phi)^{2}}$.

\subsection*{Dataset Overview}

The \emph{RODEM Jet Datasets} consist of simulated jets produced from proton-proton collisions at a center-of-mass energy of $\sqrt{s} = 13,\text{TeV}$, replicating the data-taking conditions of the LHC Run 2.
The simulations were conducted using the MadGraph5\_aMC@NLO~\cite{Alwall:2014hca} framework (v3.1.0) for hard interactions, with top-quark and $W$~boson decays modelled with MadSpin.
The mass of the top quark is set to $m_t = 173\,\text{GeV}$ for all events.
The event generation is interfaced to Pythia~\cite{Sjostrand:2014zea} (v8.243) to simulate parton shower and hadronisation.
All steps use the NNPDF2.3LO PDF set~\cite{Ball:2012cx} with $\alpha_S(m_Z) = 0.130$, as provided by the LHAPDF~\cite{Buckley:2014ana} framework.
The detector response is simulated using Delphes~\cite{deFavereau:2013fsa} (v3.4.2) with a parametrisation mimicking the response of the ATLAS detector~\cite{atlas-experiment}. 
Jets are reconstructed using the anti-$k_t$ algorithm~\cite{Cacciari:2008gp} in the FastJet implementation~\cite{Cacciari:2011ma} with a radius parameter of $R = 1.0$.

\subsection*{Simulated Processes}

The datasets include various sets of events generated from simulated proton--proton collisions at a centre-of-mass energy of $\sqrt{s} = \SI{13}{\TeV}$:
\begin{enumerate}
    \item \textbf{Light jets:}
    Simulated using QCD dijet events ($pp \to jj$), where the final-state objects, $j$, can be light quarks or gluons ($j \in [u,d,s,c,g]$, including all corresponding antiquarks).
    Their transverse momenta are restricted to $450 < \pT < \SI{1200}{\GeV}$, and their pseudorapidity is limited to $\abseta < 2.5$.
    
    \item \textbf{Jets from $\bm{W}$~bosons:}
    Simulated with $WZ$ production events ($pp \to WZ$) with $W \to jj$ and $Z \to \nu\bar{\nu}$ decays only.
    Both bosons are required to have $450 < \pT < \SI{1200}{\GeV}$.
    
    \item \textbf{Jets from top quarks:}
    Simulated with top-quark pair production events ($pp \to t\bar{t}$) with only hadronic decays of the top quarks allowed, $t \to Wb$, $W \to jj$.
    The top quarks are required to have $450 < \pT < \SI{1200}{\GeV}$.
\end{enumerate}
Additionally, semi-visible jets and jets originating from decays of heavy \ac{BSM} particles are generated as benchmark models:
\begin{enumerate}
    \item \textbf{Semi-visible jets:}
    Simulated through a dark-sector model predicting \acp{SIMP}, generated with the FeynRules package (v2.3.13), following Ref.~\cite{Bernreuther:2019pfb}.
    The processes include $pp \to Z' \to q_d \bar{q}_d$, where $q_d$ denotes a dark-sector quark, and $Z'$ is a generic spin-1 mediator with vector couplings to both \ac{SM} and dark-sector quarks.
    Each dark quark is required to have $600 < \pT < \SI{1600}{\GeV}$.
    
    \item \textbf{Resonant Higgs boson production:}
    Simulated using a type-II \ac{2HDM} generated with the FeynRules package~\cite{Alloul_2014} (v2.3.24).
    The processes include $pp \to H^0 \to h^{+}h^{-}$, where $H$ denotes a heavy neutral Higgs scalar and $h^{\pm}$ a lighter charged Higgs scalar.
    Production via gluon-gluon fusion and $b\bar{b}$ annihilation is considered.
    Two types of decays are simulated: $h \to jj$ and $h \to tb$, leading to two-prong and four-prong jet substructures, respectively.
    The top quarks in the latter case are subsequently decayed as in the \ac{SM} $t\bar{t}$ simulation.
    Sets of \num{100}k events were generated for both production modes, both decay modes, and each of the following $(m_H, m_h)$ mass grid points: $(1300, 200)$, $(1300, 250)$, $(1350, 300)$, $(1700, 250)$, $(1700, 300)$, $(1800, 400)$, $(2100, 300)$, $(2150, 400)$, $(2250, 500)\,\si{\GeV}$.
    
\end{enumerate}

\subsection*{Event Selection}

Reconstructed events are selected by requiring a jet with large transverse momentum, $\pT > \SI{450}{\GeV}$.
For all simulations except $WZ$, a second jet with $\pT > \SI{200}{\GeV}$ is required.
No other event selection criteria are imposed; additional jets are not vetoed.
The numbers of events remaining after selection are listed in \cref{tab:nevents}.

\begin{table}[tp]
    \caption{
        Number of events in the datasets after event selection.
        All simulations except $WZ$ contain two large-radius jets.
        The number of \ac{2HDM} events varies depending on the generated $m_H$ and $m_h$ values.
        The right-most column indicates the train/test/validation split the datasets are provided with.
        No split is performed for the \ac{2HDM} datasets due to the small sample sizes.
        }
    \label{tab:nevents}
    \centering
    \begin{tabular}{lS[table-format=9,round-mode=figures,round-precision=3] c}
        \toprule
         Process & {\# events} & {train/test/val split} \\
         \midrule
         QCD dijet    & 9559033 & $90 : 5 : 5$ \\
         $t\bar{t}$   & 14987533 & $90 : 5 : 5$ \\
         $WZ$         & 14073844 & $90 : 5 : 5$ \\
         Semi-visible & 908159 & $90 : 5 : 5$ \\
         \ac{2HDM}    & \leq 90389 & None \\
        \bottomrule
    \end{tabular}
\end{table}

\subsection*{Dataset Structure and Content}

\begin{figure*}[hp]
    \centering
    \includegraphics[width=0.32\textwidth, trim=0 10 0 10, clip=true]{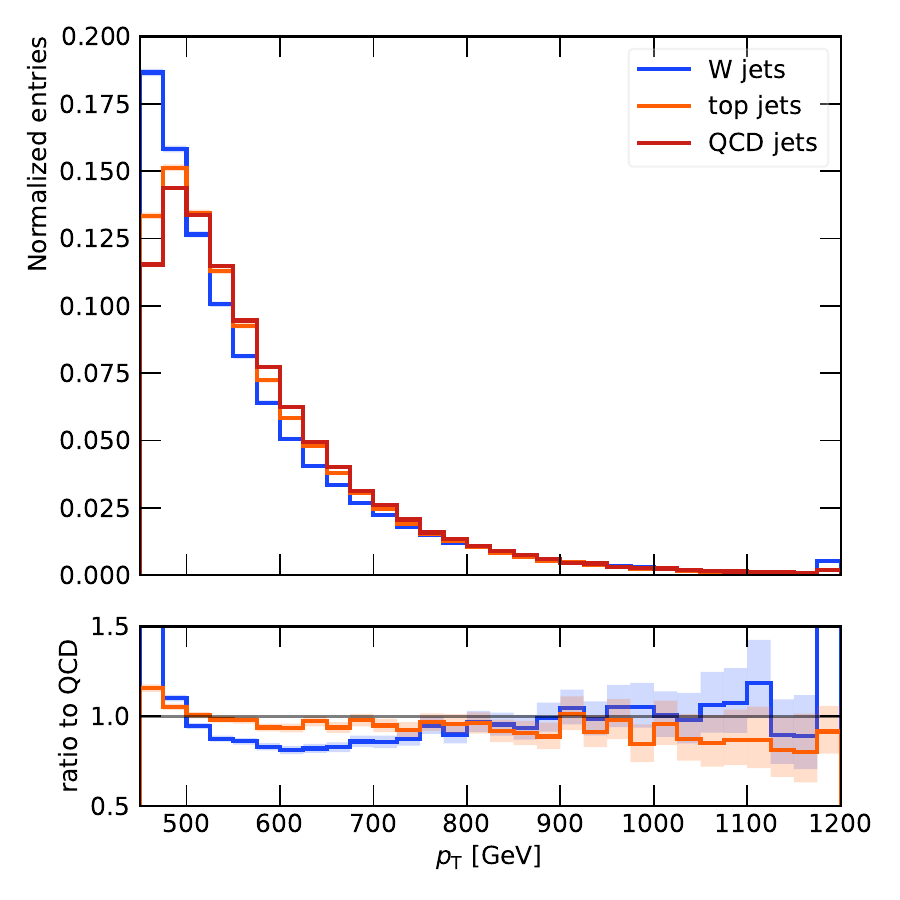}%
    \includegraphics[width=0.32\textwidth, trim=0 10 0 10, clip=true]{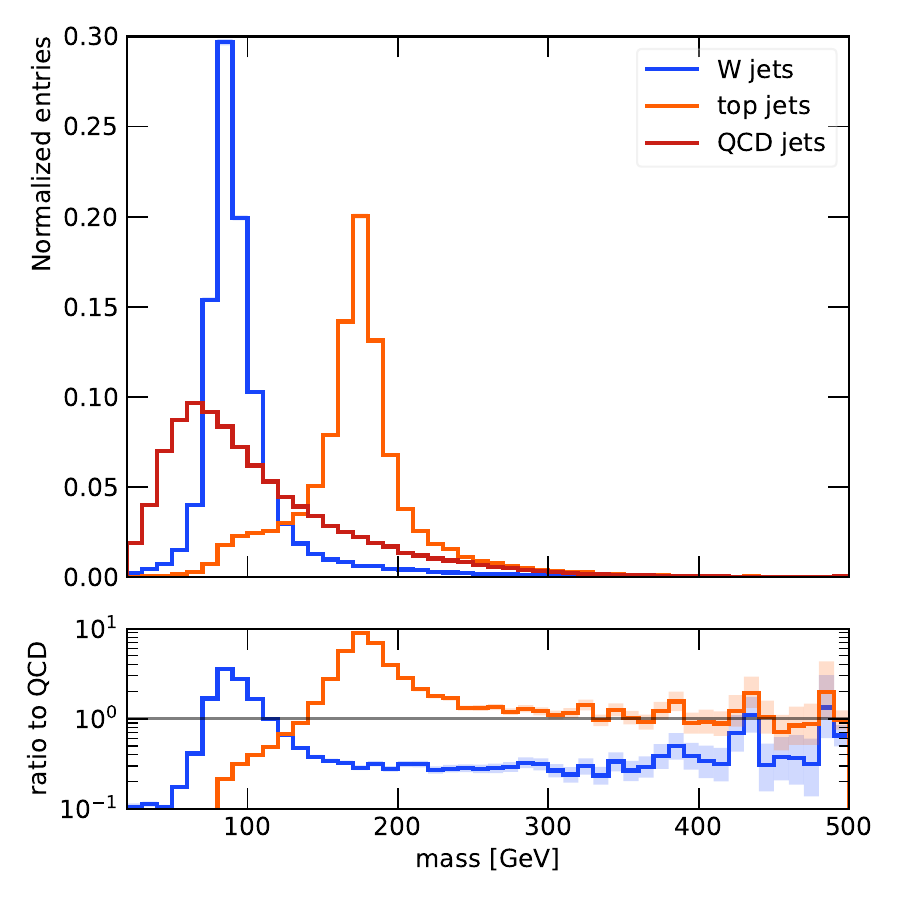}%
    \includegraphics[width=0.32\textwidth, trim=0 10 0 10, clip=true]{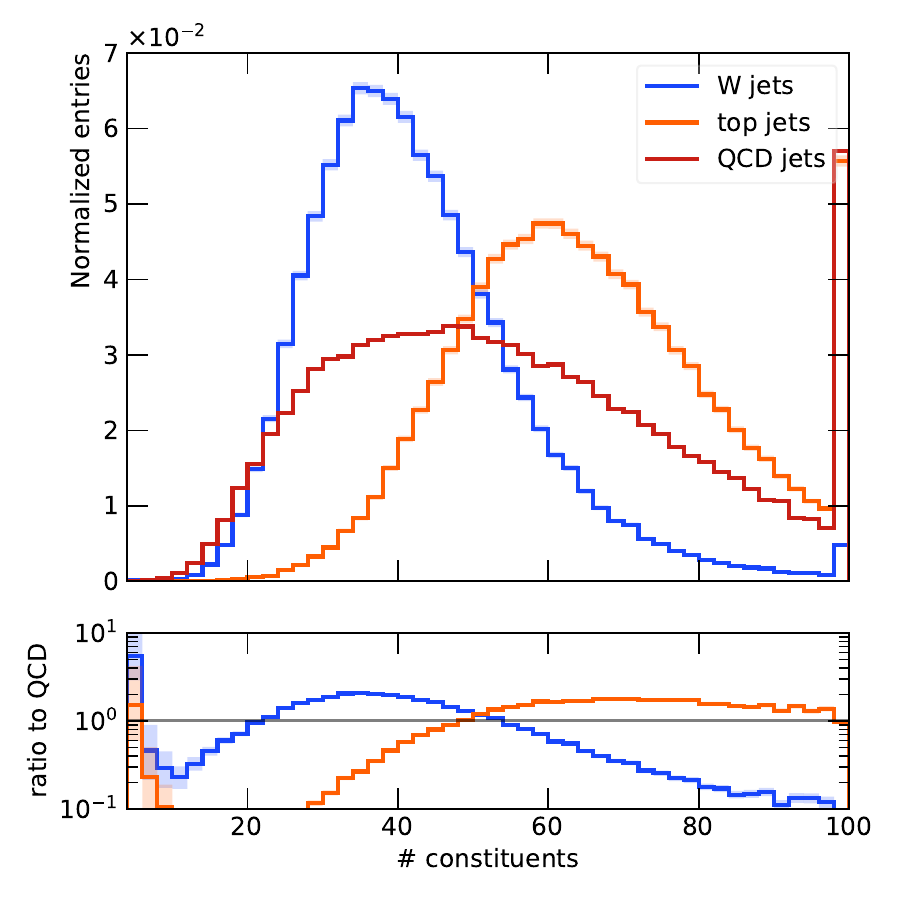}\\
    \includegraphics[width=0.32\textwidth, trim=0 10 0 10, clip=true]{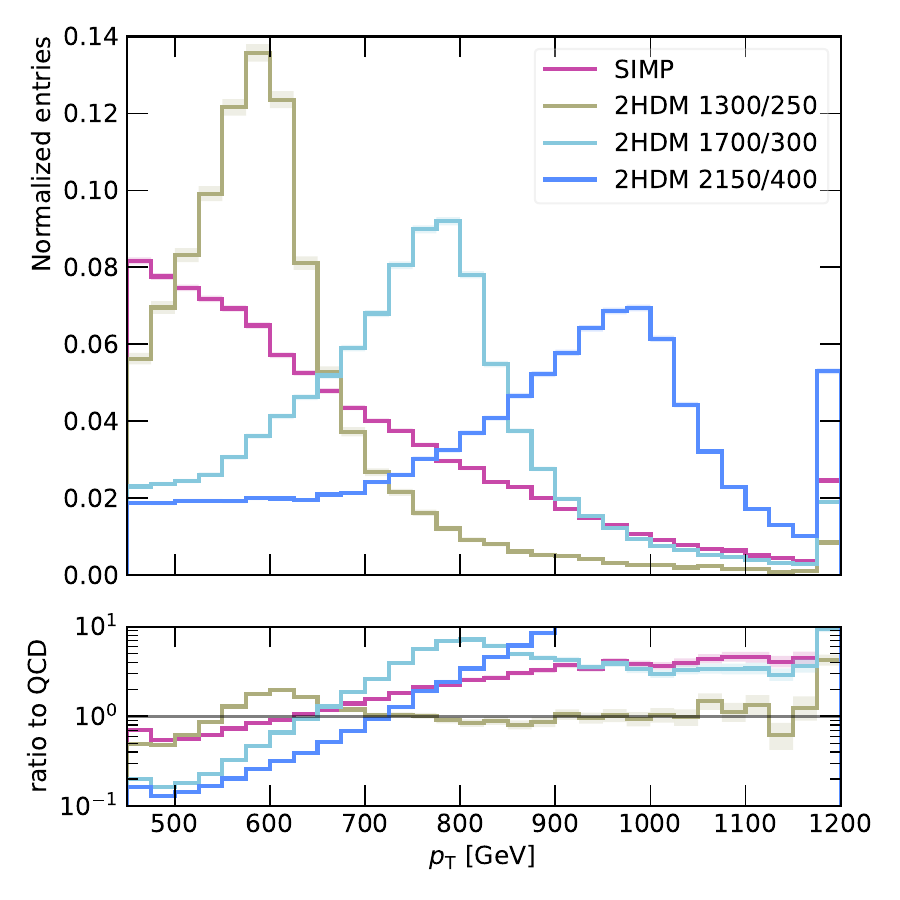}%
    \includegraphics[width=0.32\textwidth, trim=0 10 0 10, clip=true]{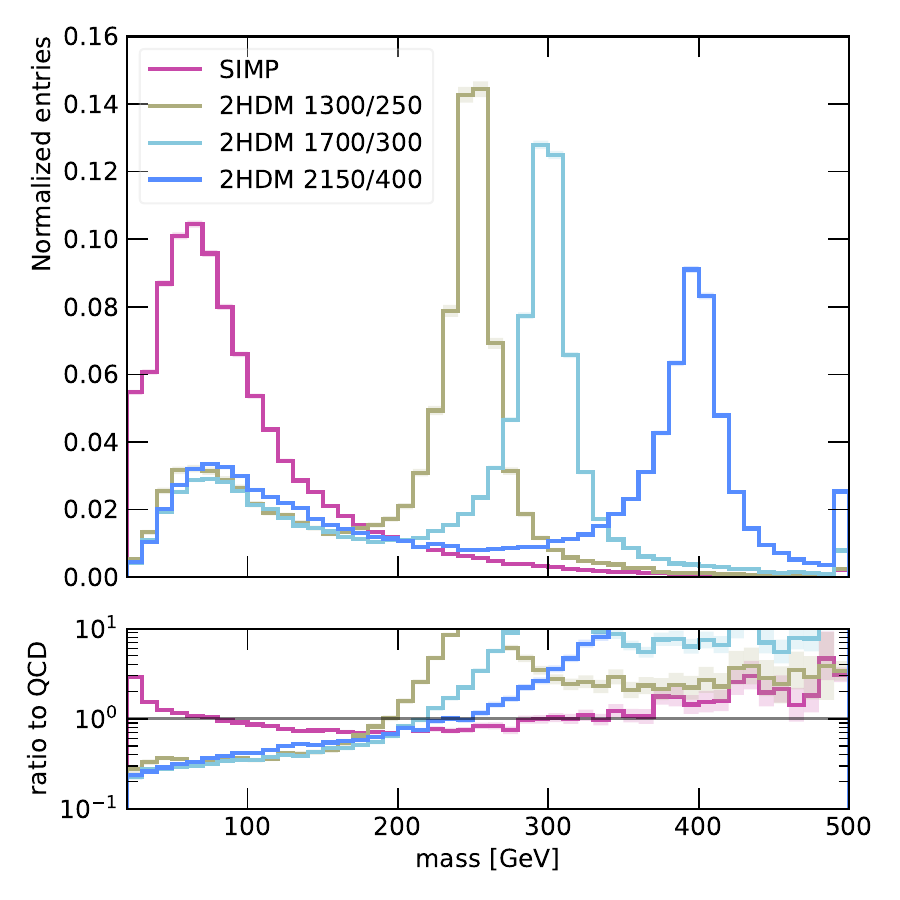}%
    \includegraphics[width=0.32\textwidth, trim=0 10 0 10, clip=true]{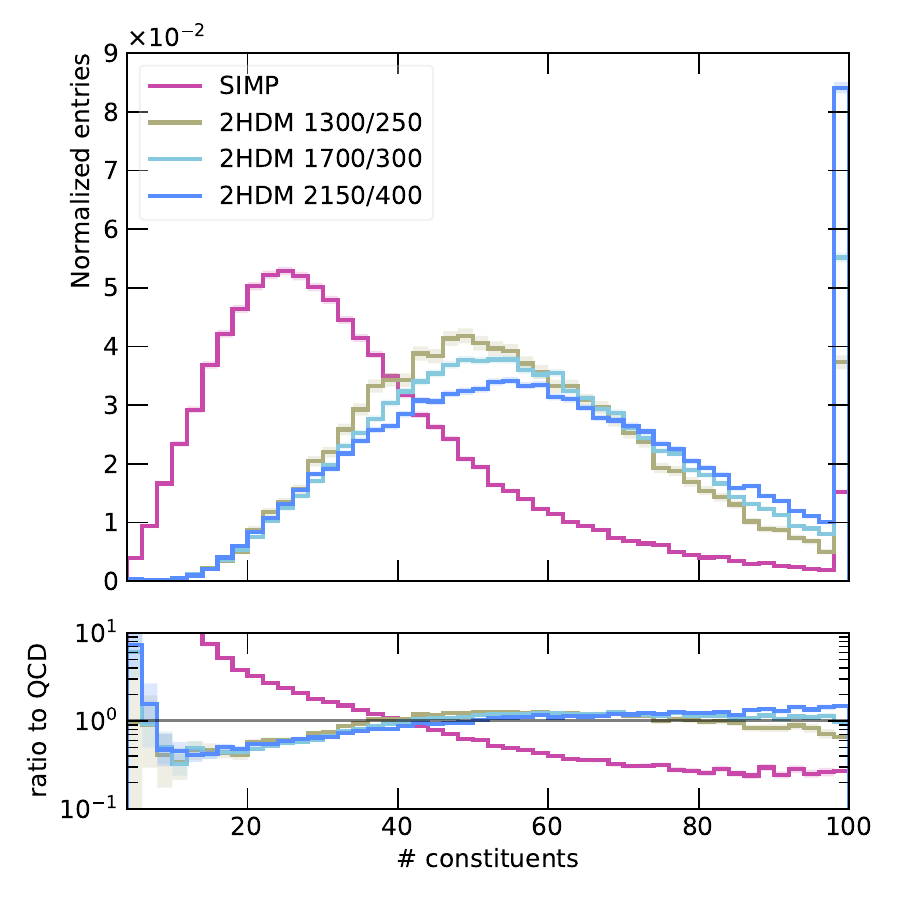}\\
    \includegraphics[width=0.32\textwidth, trim=0 10 0 10, clip=true]{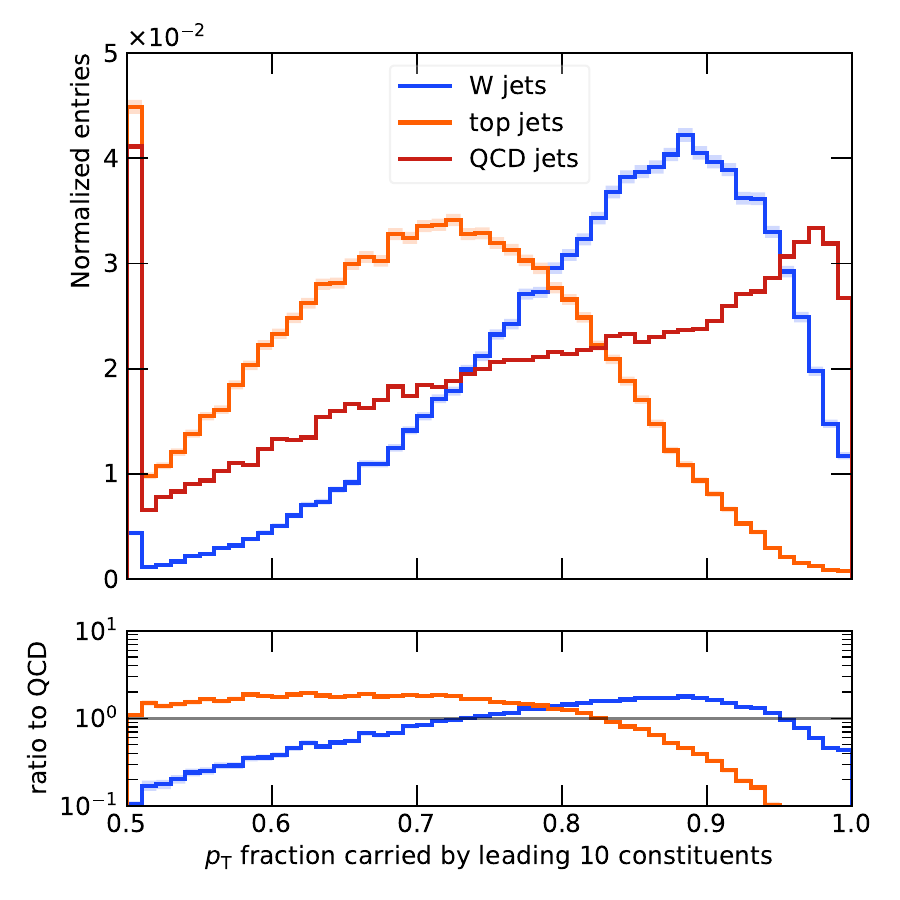}%
    \includegraphics[width=0.32\textwidth, trim=0 10 0 10, clip=true]{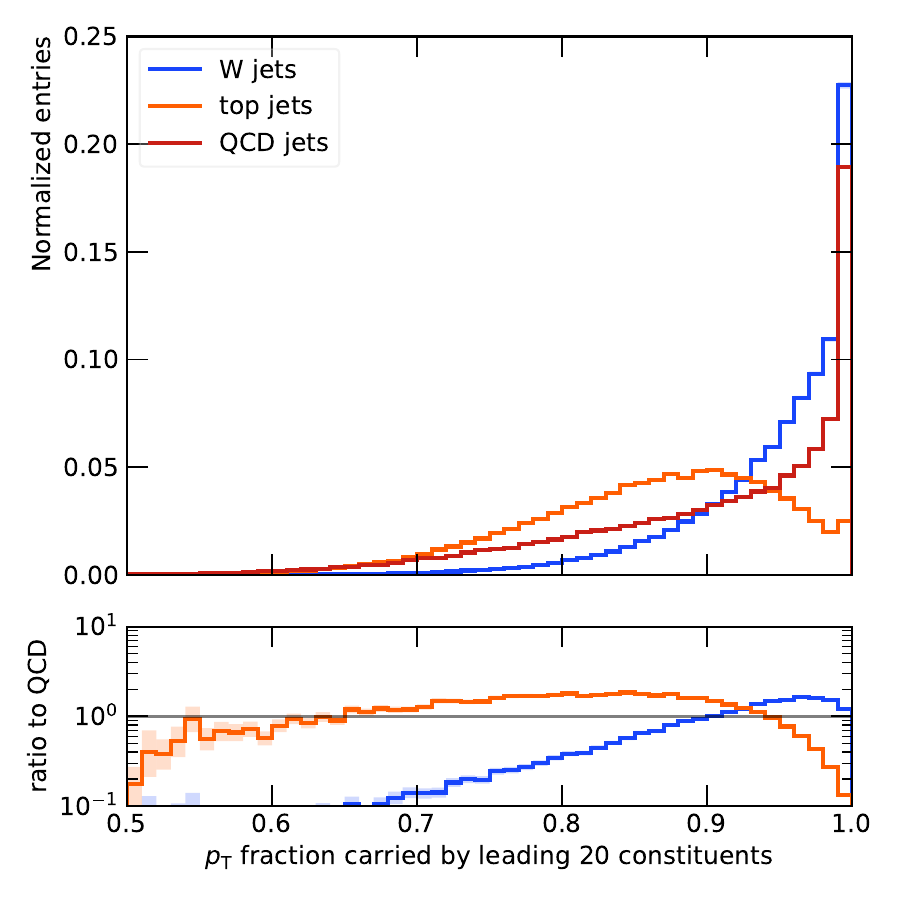}%
    \includegraphics[width=0.32\textwidth, trim=0 10 0 10, clip=true]{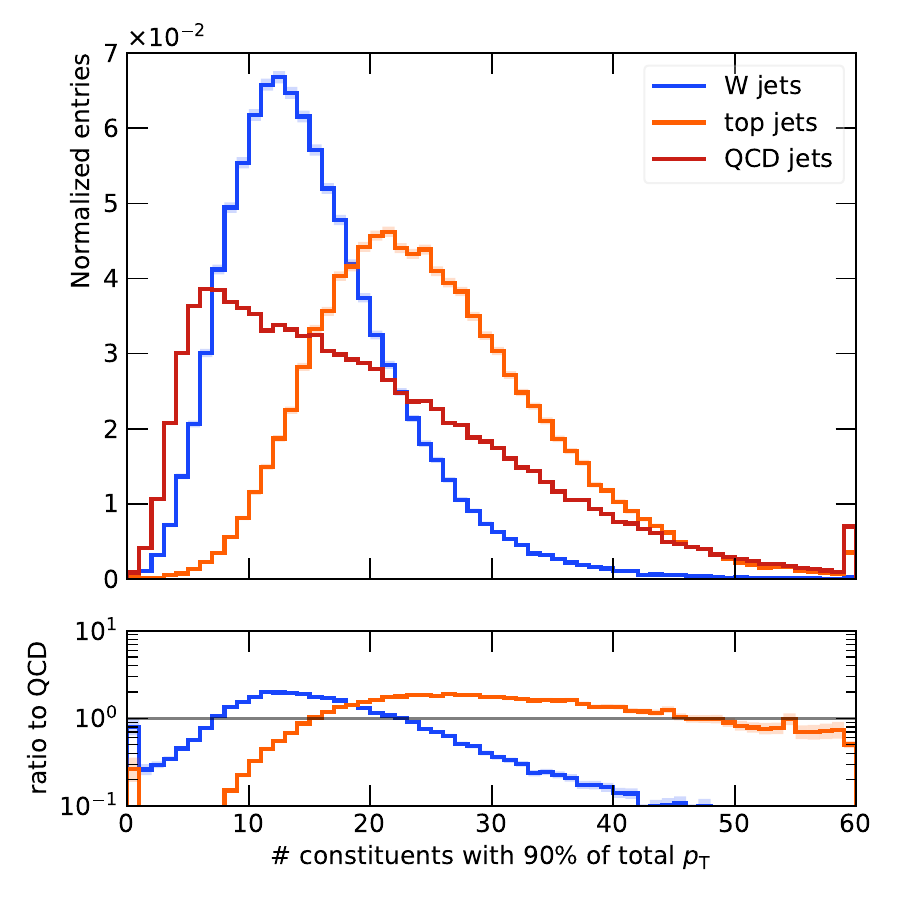}\\
    \includegraphics[width=0.32\textwidth, trim=0 10 0 10, clip=true]{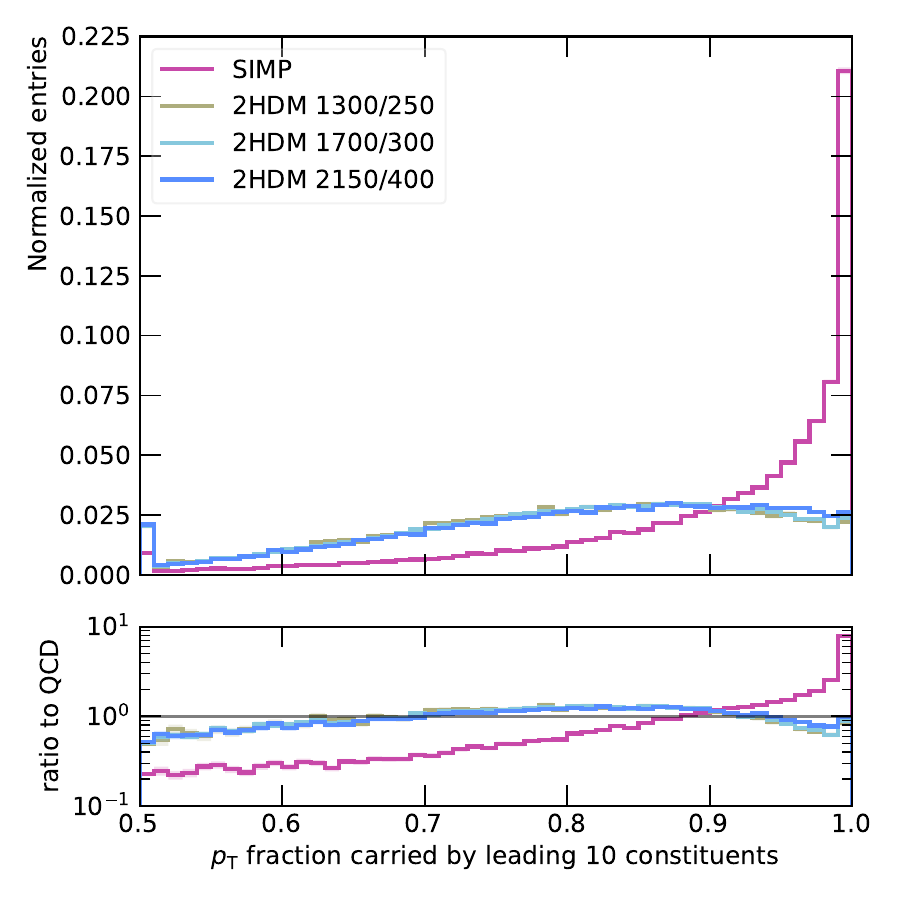}%
    \includegraphics[width=0.32\textwidth, trim=0 10 0 10, clip=true]{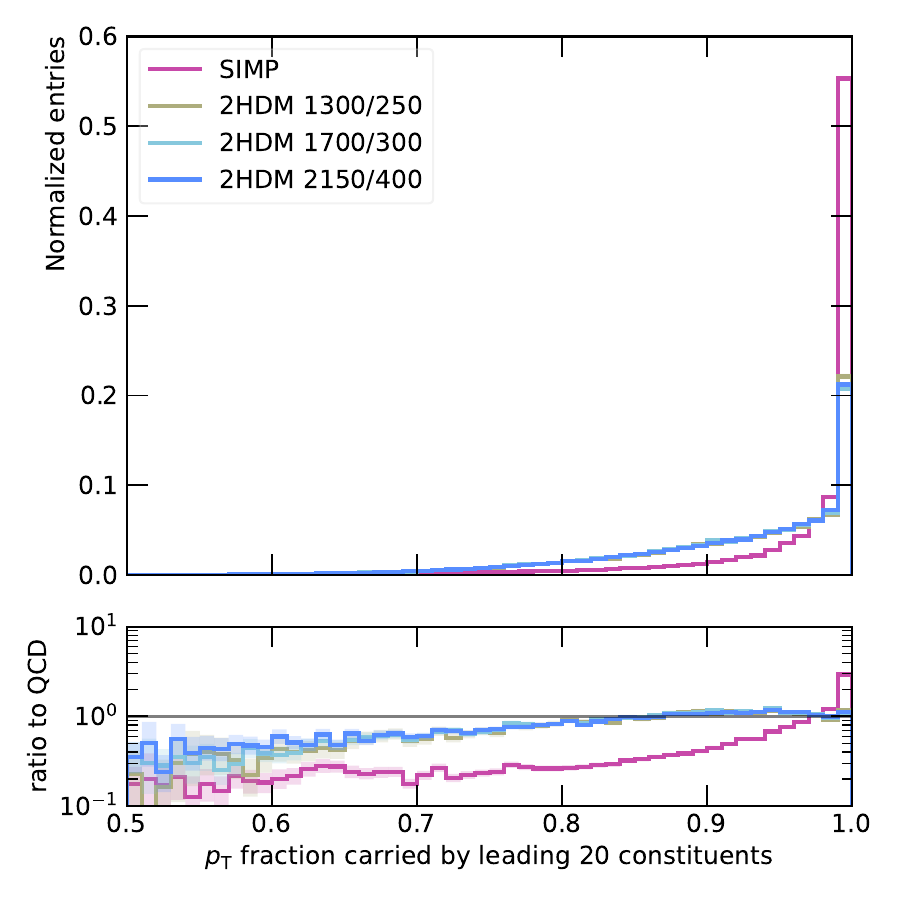}%
    \includegraphics[width=0.32\textwidth, trim=0 10 0 10, clip=true]{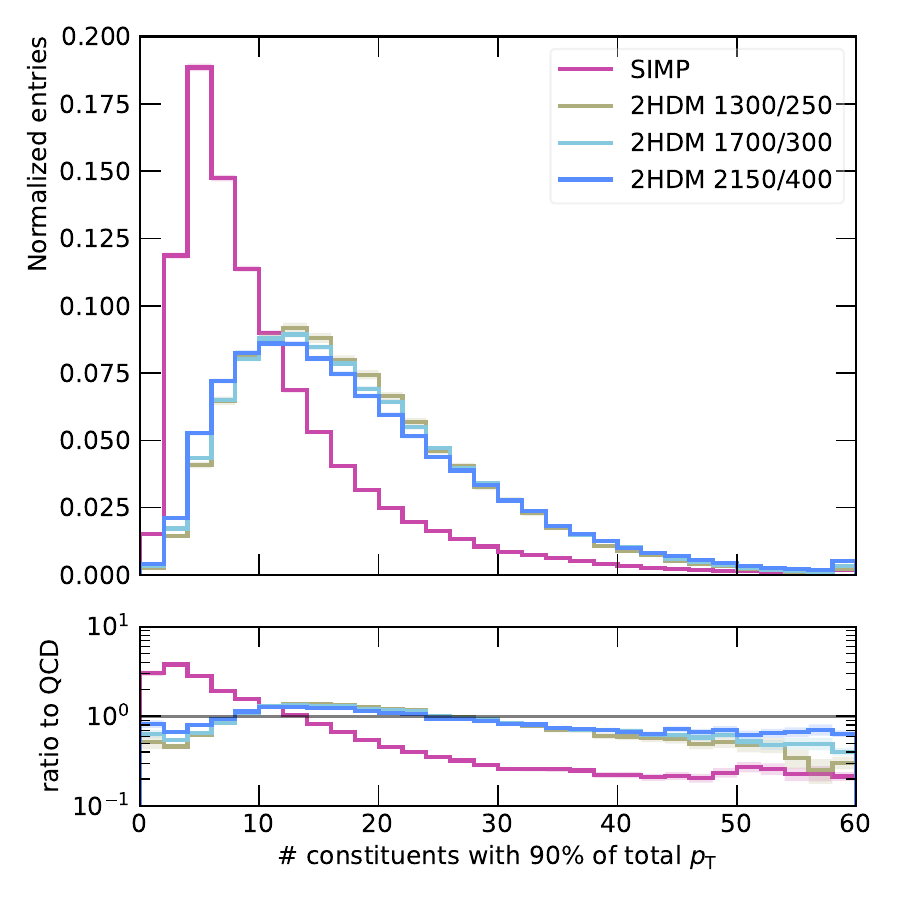}
    \caption{%
        Basic kinematics and constituent-level information for the \ac{SM} samples and four exemplary \ac{BSM} samples (SIMP and \ac{2HDM} $gg \to H \to hh \to jjjj$ with three different mass settings).
        Top two rows, left to right: jet transverse momentum, mass and number of constituents.
        Bottom two rows, left to right: \pT fraction carried by the 10 leading constituents, \pT fraction carried by the 20 leading constituents, number of constituents carrying 90\% of the total \pT.
    }
    \label{fig:kins_cnsts}
\end{figure*}

The datasets are provided in the \texttt{.hdf5} format.
To facilitate their use for \ac{ML} applications, the large simulated jet datasets are stored using a train\,:\,validation\,:\,test split of approximately 90\,:\,5\,:\,5 (see \cref{tab:nevents} for details).
When opening the files, each file contains a group named \texttt{objects/jets}, which in turn contains four datasets:
\begin{itemize}
    \item \texttt{jet1\_obs}: Jet-level observables for the leading jet in \pT. The dimensions of the dataset are $(N, 11)$ for $N$~events, with columns representing \pT, $\eta$, $\phi$ and mass of the jet, and seven substructure metrics, see the following section.
    
    \item \texttt{jet1\_cnsts}: Constituents of the leading jet in \pT, sorted by \pT. The dimensions are $(N, 100, 7)$.
    If a jet has more than 100 constituents, only the 100 leading in \pT are stored; if fewer, the dataset is zero-padded.
    The seven columns include the constituents' \pT, $\eta$, $\phi$, mass, electric charge, and transverse and longitudinal impact parameters ($d_0$, $d_z$).
    The last three are only filled for electrically charged constituents and are zero-padded otherwise.
    
    \item \texttt{jet2\_obs}: Jet-level observables for the jet subleading in \pT.
    
    \item \texttt{jet2\_cnsts}: Constituents of the jet subleading in \pT.
\end{itemize}
Note that the last two datasets are not present for $WZ$~events.
\Cref{fig:kins_cnsts} shows distributions of jet kinematics and constituent-level information for the \ac{SM} datasets and four exemplary \ac{BSM} datasets (SIMP and \ac{2HDM} $gg \to H \to hh \to jjjj$ with three different mass settings).

\subsection*{Substructure Metrics}

\begin{figure*}[hp]
    \centering
    \includegraphics[width=0.32\textwidth, trim=0 10 0 10, clip=true]{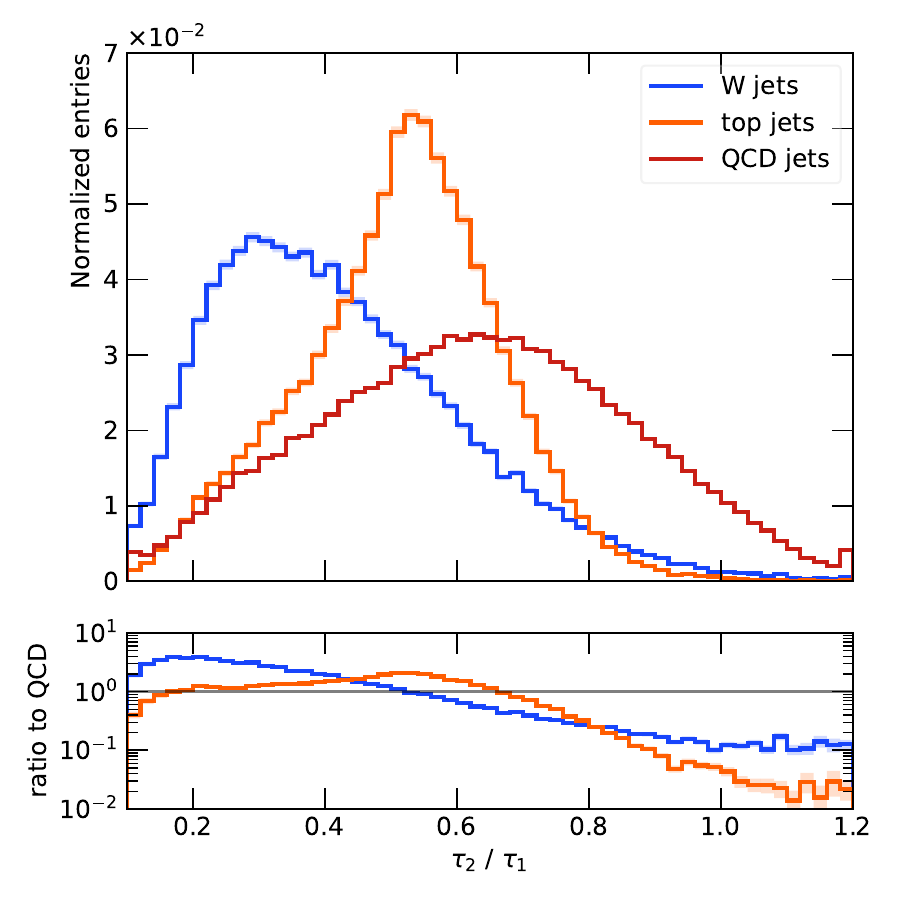}%
    \includegraphics[width=0.32\textwidth, trim=0 10 0 10, clip=true]{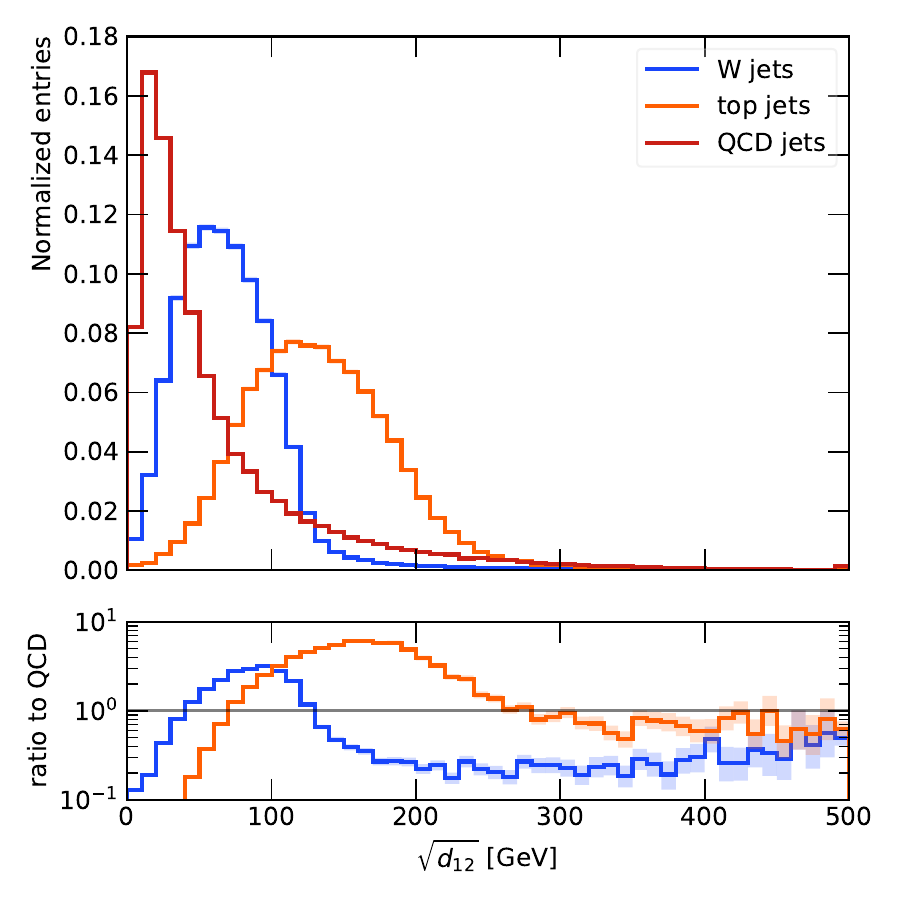}%
    \includegraphics[width=0.32\textwidth, trim=0 10 0 10, clip=true]{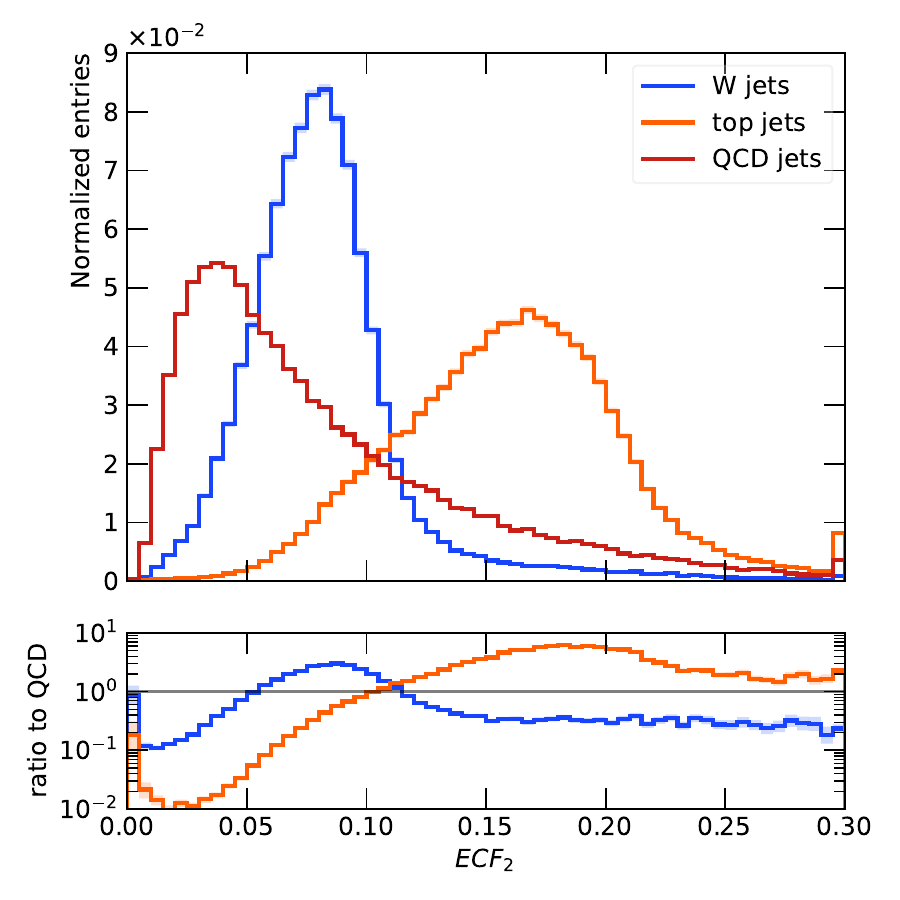}\\
    \includegraphics[width=0.32\textwidth, trim=0 10 0 10, clip=true]{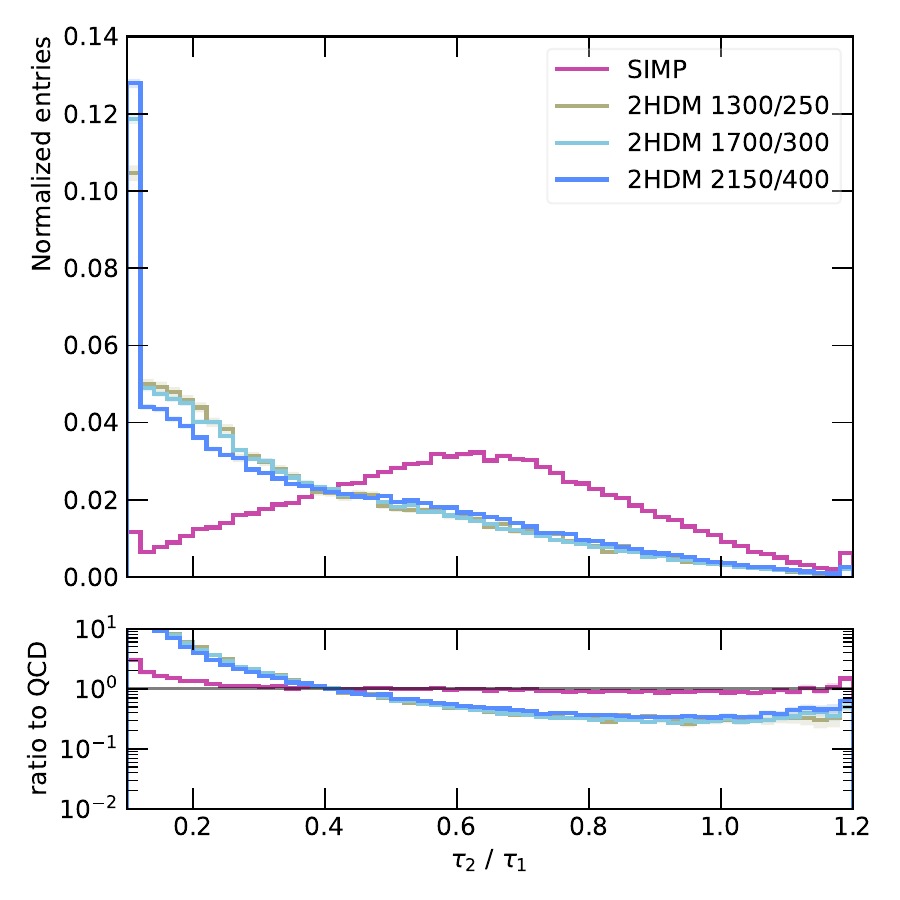}%
    \includegraphics[width=0.32\textwidth, trim=0 10 0 10, clip=true]{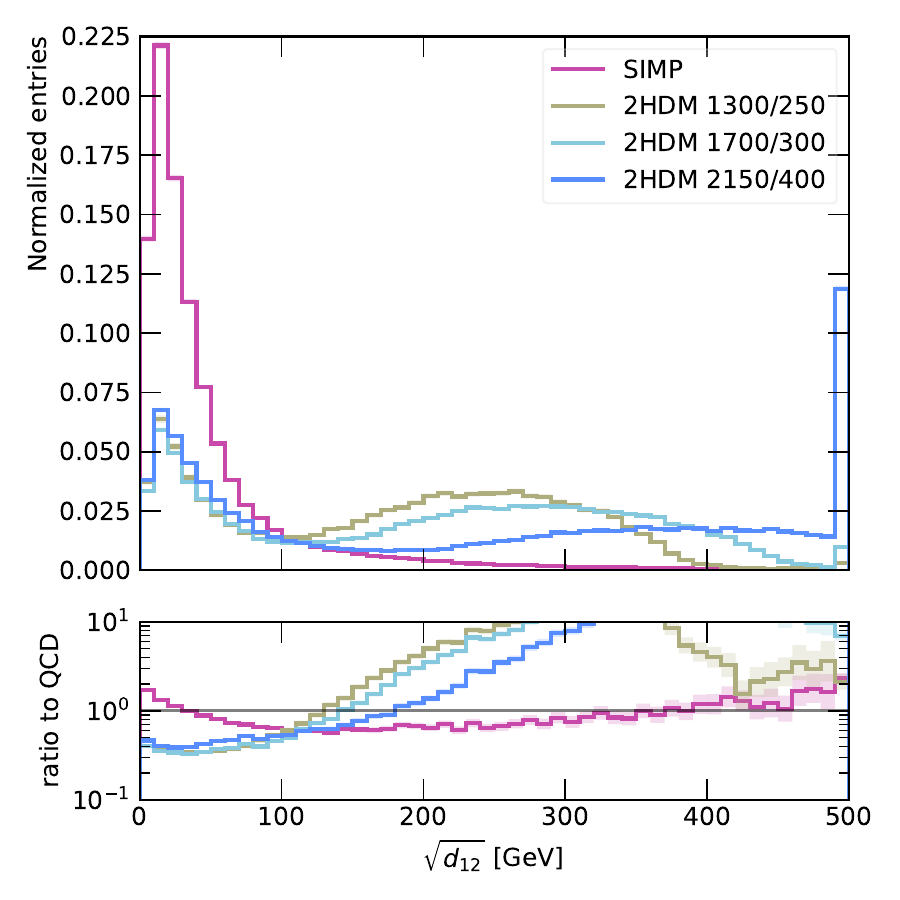}%
    \includegraphics[width=0.32\textwidth, trim=0 10 0 10, clip=true]{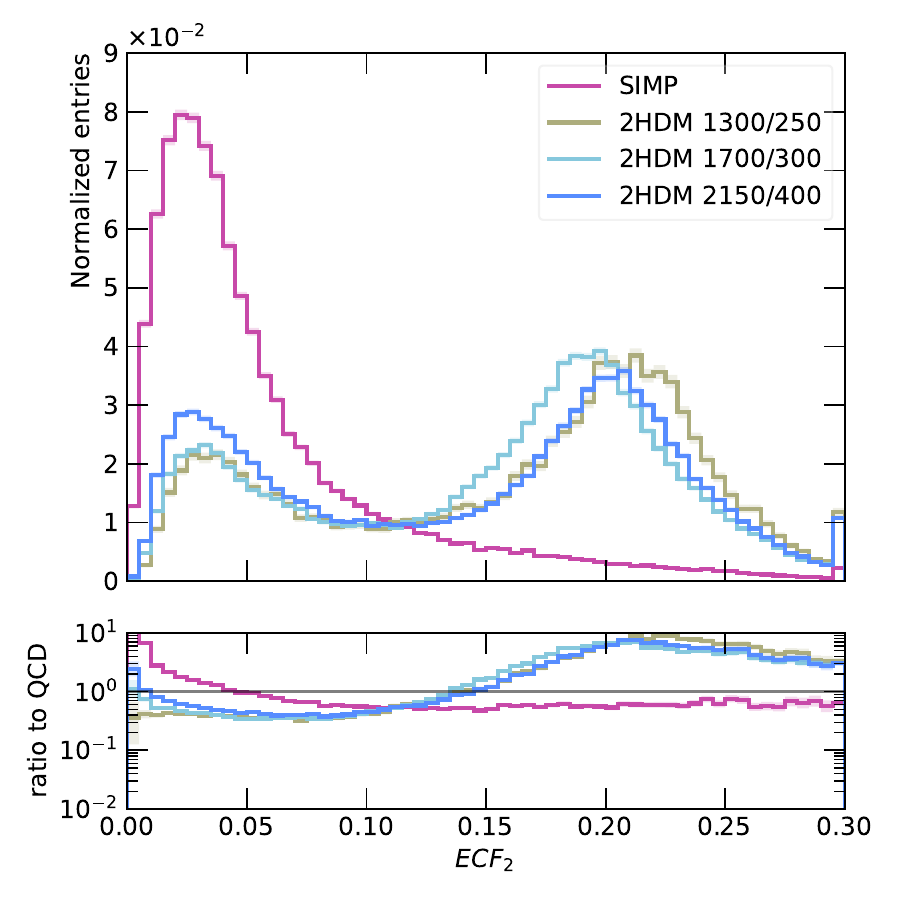}\\
    \includegraphics[width=0.32\textwidth, trim=0 10 0 10, clip=true]{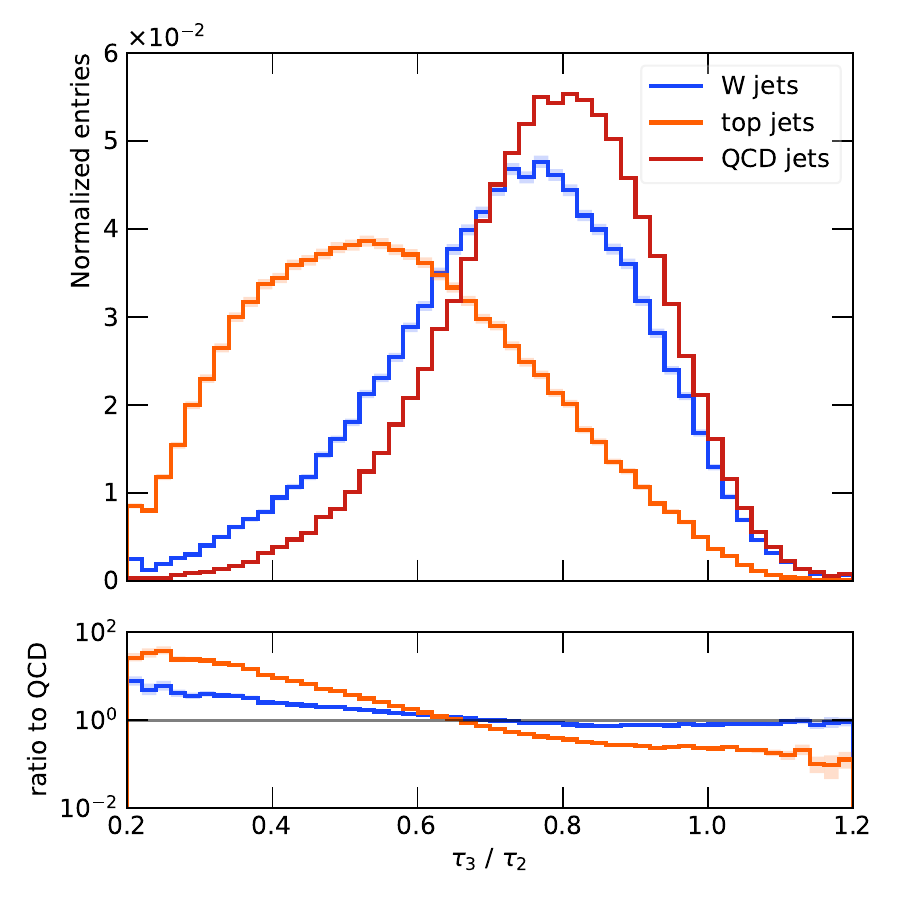}%
    \includegraphics[width=0.32\textwidth, trim=0 10 0 10, clip=true]{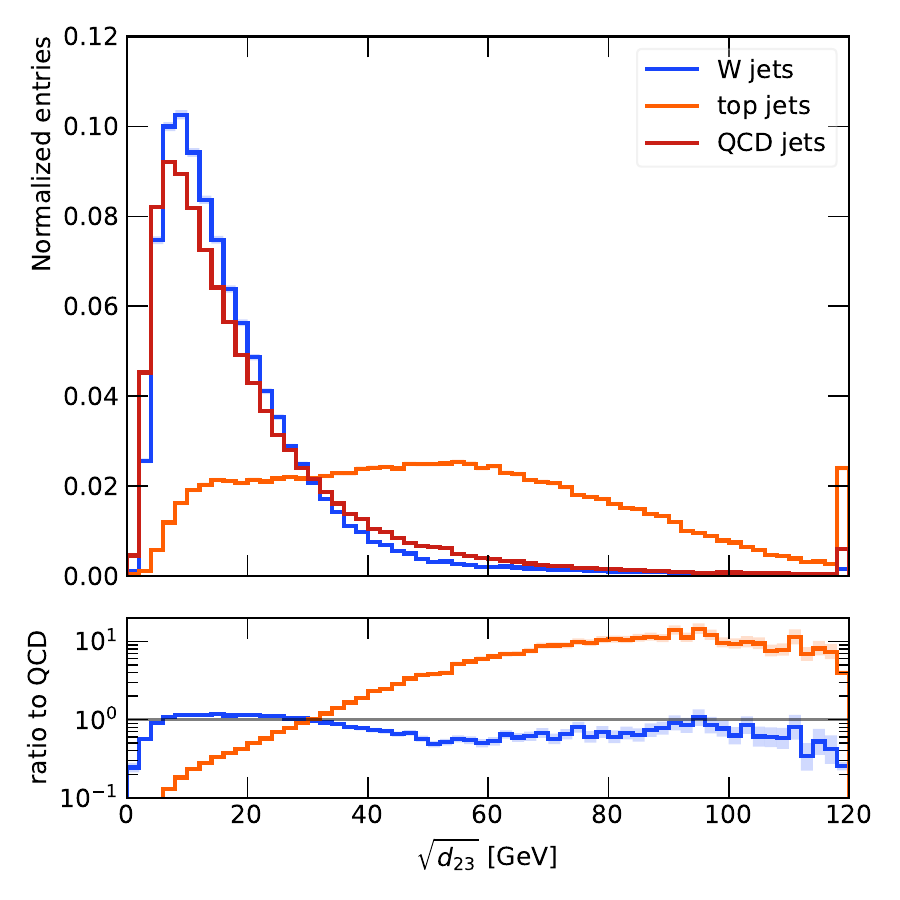}%
    \includegraphics[width=0.32\textwidth, trim=0 10 0 10, clip=true]{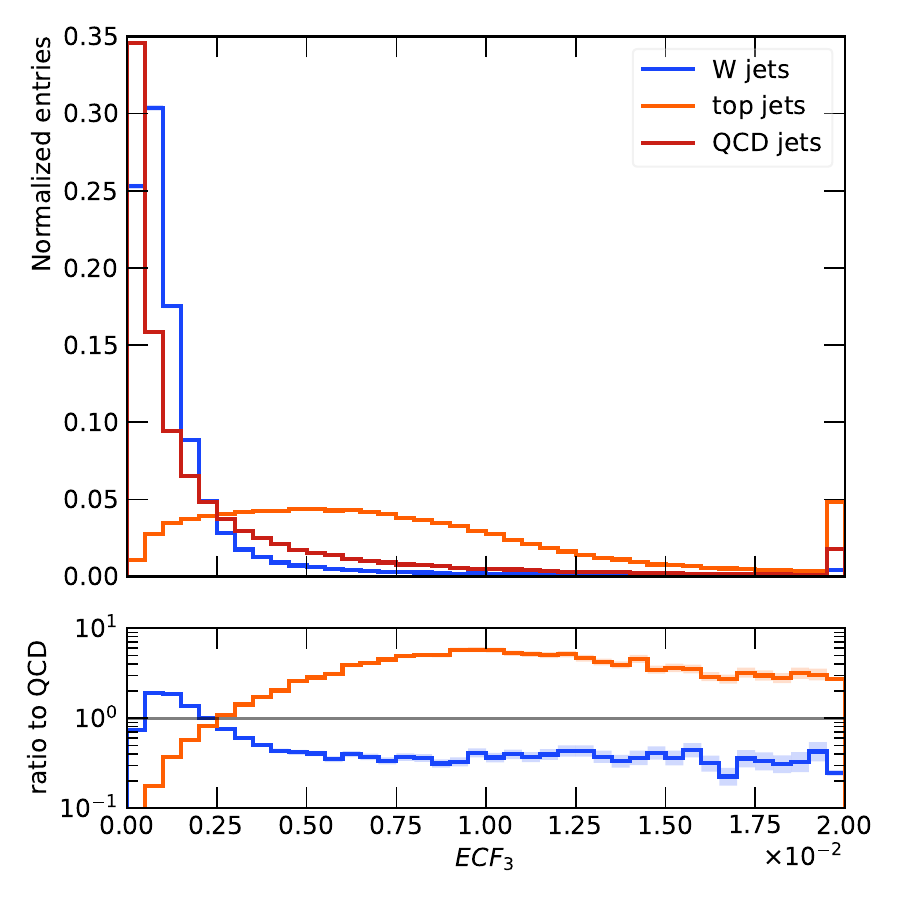}\\
    \includegraphics[width=0.32\textwidth, trim=0 10 0 10, clip=true]{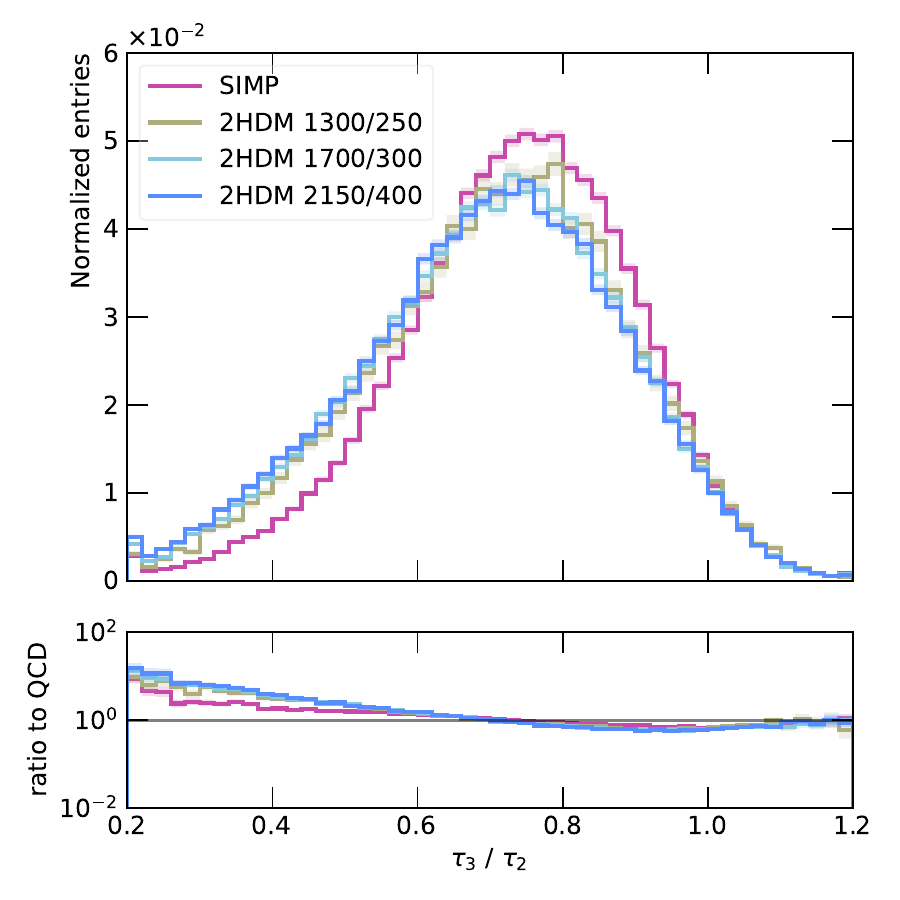}%
    \includegraphics[width=0.32\textwidth, trim=0 10 0 10, clip=true]{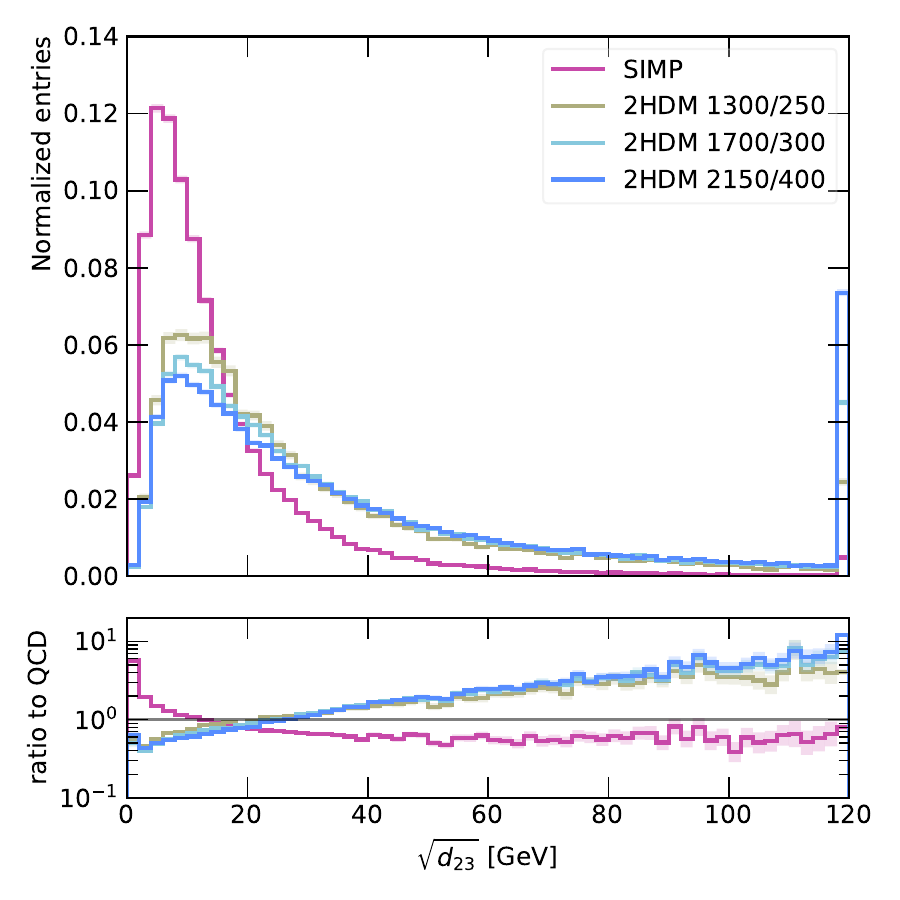}%
    \includegraphics[width=0.32\textwidth, trim=0 10 0 10, clip=true]{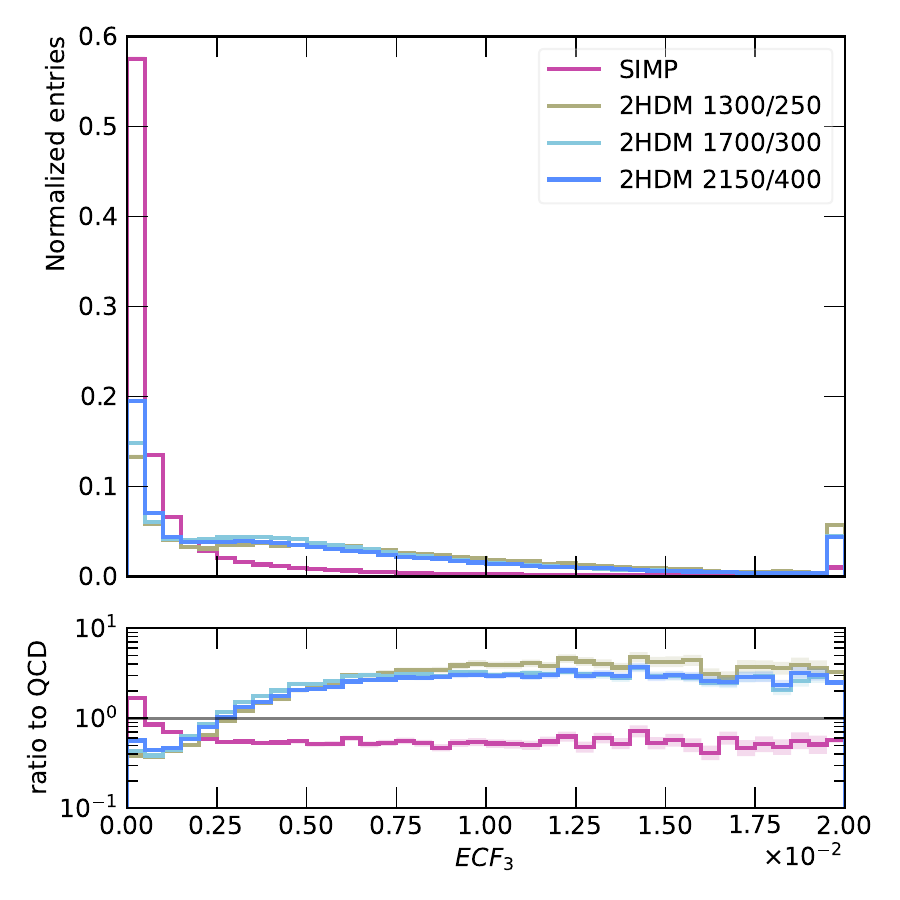}
    \caption{
        Substructure metrics for the jets in the \ac{SM} datasets and four exemplary \ac{BSM} datasets (SIMP and \ac{2HDM} $gg \to H \to hh \to jjjj$ with three different mass settings).
        Top two rows, left to right: $N$-subjettiness ratio $\tau_2 / \tau_1$, splitting scale $\sqrt{d_{12}}$, and energy correlation functions $\mathit{ECF}_2$.
        Bottom two rows, left to right: $\tau_3 / \tau_2$,  $\sqrt{d_{23}}$, and $\mathit{ECF}_3$.
    }
    \label{fig:substructure}
\end{figure*}

In addition to the raw constituent kinematic, the jet-level observable datasets store seven substructure metrics that capture the internal structure of the jet, according to the following definitions:
\begin{enumerate}
\item \textbf{$\bm{N}$-subjettiness:} 
For these metrics, the jet constituents are re-clustered using $N$-exclusive $k_t$ clustering~\cite{Catani:1993hr,Salam:2010nqg} into $N$~subjets.
The $N$-subjettiness values~\cite{nsubjettiness} are computed as
\begin{equation}
    \tau_N = \frac{1}{d_0 R_0} \sum_{i} \pTindex{i} \min\{\Delta R_{1,i}, \Delta R_{2,i}, \ldots, \Delta R_{N,i}\}
\end{equation}
where $\Delta R_{k,i}$ is the distance between the $k$-th subjet axis and the $i$-th constituent, 
$d_0 = \sum_{i} \pTindex{i}$ is a normalisation factor to convert $\tau_N$ into a dimensionless quantity,
and $R_0=1.0$ is the radius parameter of the original anti-$k_t$ clustering.
In the datasets, we store \tauone, \tautwo, and \tauthree.

\item \textbf{Exclusive splitting scales}:
Comparing exclusive $k_t$ clustering into $N$ and $N+1$ subjets, one of the $N$ subjets splits in two. 
The exclusive splitting scale $\sqrt{d_{N,N+1}}$ measures the scale at which this additional splitting occurs:
\begin{equation}
    \sqrt{d_{N,N+1}} = \sqrt{1.5} \cdot \min\{\pTindex{i}, \pTindex{j}\} \cdot \frac{\Delta R_{i,j}}{R_0}
\end{equation}
where $i$ and $j$ are the two subjets resulting from the additional split.
In the datasets, we store \donetwo and \dtwothree.

\item \textbf{Energy correlation functions:} For these metrics~\cite{ecf1}, we store the 2-point and 3-point energy correlation functions, \ecftwo and \ecfthree.
These are calculated as:
\begin{align}
    \ecftwo &= \frac{1}{d_0^2} \sum_{i<j} \pTindex{i}\; \pTindex{j} \left( \Delta R_{ij} \right)^\beta \\
    \ecfthree &= \frac{1}{d_0^3} \sum_{i<j<k} \pTindex{i}\; \pTindex{j}\; \pTindex{k} \cdot \left( \Delta R_{ij} \Delta R_{ik} \Delta R_{jk} \right)^\beta
\end{align}
where the $d_0$ factors convert them into dimensionless quantities.
For this dataset, $\beta = 1$.
\end{enumerate}
Distributions of the substructure metrics for the \ac{SM} datasets and four exemplary \ac{BSM} datasets are shown in \cref{fig:substructure}.

\section*{Acknowledgements}

We sincerely thank Dr. Matthias J. Schlaffer for his valuable contributions to fruitful discussions and the establishment of our 2HDM sample production.

The authors would like to acknowledge funding through the SNSF Sinergia grant CRSII5\_193716 ``Robust Deep Density Models for High-Energy Particle Physics and Solar Flare Analysis (RODEM)'', the SNSF project grant 200020\_212127 ``At the two upgrade frontiers: machine learning and the ITk Pixel detector'', and the SNSF project grant 200020\_181984 ``Exploiting LHC data with machine learning and preparations for HL-LHC''.
They would also like to acknowledge individual funding acquired through the Feodor Lynen Research Fellowship from the Alexander von Humboldt foundation.
The computations were performed at the University of Geneva using the Baobab HPC service.

\printbibliography[title=References]

\end{document}